\documentclass[twocolumn]{book}
\usepackage[dvips]{graphicx,color}
\usepackage{makeidx,universe}

\makeauthorindex

\BookTitle{Proceedings of the XXIX PHYSICS IN COLLISION}

\CopyRight{\copyright 2009 by Universal Academy Press, Inc.}

\begin{document} 

\pagenumbering{arabic}

\chapter{%
{\LARGE \sf
Higgs searches} \\
{\normalsize \bf 
Krisztian Peters} \\
{\small \it \vspace{-.5\baselineskip}
School of Physics and Astronomy, The University of Manchester,
Oxford Road, Manchester M13 9PL, United Kingdom}
}


\AuthorContents{K.\ Peters}

\AuthorIndex{Peters}{K.}

  \baselineskip=10pt 
  \parindent=10pt    

\section*{Abstract} 

We present the status and prospects of Higgs searches at the Tevatron and the LHC. Results from the 
Tevatron are using up to 
5~fb$^{-1}$ of data collected with the CDF and D0 detectors.
The major contributing processes include associated production ($WH\to l\nu bb$, $ZH\to\nu\nu bb$, $ZH
\to ll bb$) and gluon fusion ($gg\to H\to WW^{(*)}$). 
Improvements across the full mass range resulting from the larger data sets, improved analyses techniques 
and increased signal acceptance are discussed. Recent results exclude the SM Higgs boson in a mass 
range of $160 < m_{H}< 170$ GeV. Searches for the neutral MSSM
Higgs boson in the region $90 < m_{A} < 200$ GeV exclude $\tan \beta$ values down to 30 for several 
benchmark scenarios. 

\section{Introduction} 

In the Standard Model (SM) the fermions and vector bosons acquire their masses 
through the Higgs mechanism. It predicts the existence of a heavy scalar boson, 
the Higgs boson, with a mass that cannot be predicted by the SM.
The Higgs particle has been searched for for decades but it is the 
only particle of the Standard Model which has yet to be directly observed. 

Such searches have been performed at the LEP experiments 
where the principal production channel is Higgs production associated with a 
$Z$ boson. The kinematic limit for these searches was the collider energy modulo the $Z$ mass which gives 
approximately 115 GeV. The combined searches of the four LEP experiments  
resulted in an exclusion of SM Higgs bosons with a mass lower than 
114.4 GeV at the 95\% confidence level (C.L.) \cite{lep}. This is still the lower bound for the mass of the
SM Higgs boson from direct searches up to date.

An upper bound
on the Higgs mass can be obtained by global electroweak fits. The $\Delta\chi^{2}$ with respect to the 
minimum of these fits is displayed in Fig.~1. 
Here the $W$ mass and the top quark mass play an important role because they are directly related to the 
Higgs mass by radiative corrections. 
New precision measurements of the $W$ mass
\cite{top} and the top mass \cite{W} from the Tevatron favor a light SM
Higgs boson and yield an upper value of 157 GeV at 95\% C.L. (or 186
GeV if the LEP2 limit is included) \cite{ewfit}.

The results from direct searches at LEP and from the global electroweak fits 
indicate that if the SM is correct the Higgs mass is likely to be somewhere between 
100 and 200 GeV. This region is within reach 
of the Tevatron and LHC colliders, thus searching for the Higgs is of highest priority in their physics program. 

Several theories beyond the SM predict the existence of one or more than one Higgs boson as well. 
The Minimal Supersymmetric Standard Model (MSSM) predicts two Higgs
doublets leading to five physical Higgs bosons. Constraints from the combined result of the four LEP 
experiments to the MSSM parameter space \cite{lepmssm} are being extended by the Tevatron experiments 
with increasing 
integrated luminosities.

\begin{figure}[t]
  \begin{center}
    \includegraphics[height=17pc]{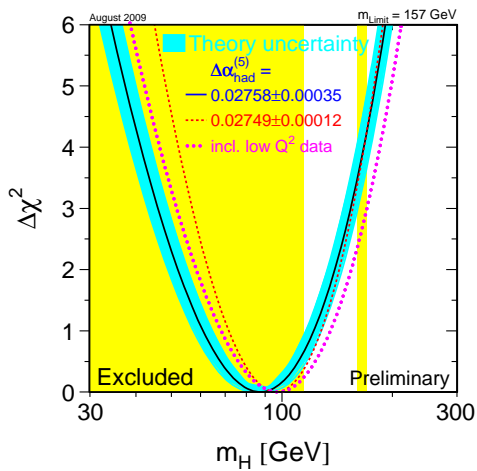}
  \end{center}
  \vspace{-1pc}
  \caption{$\Delta\chi^{2}$ with respect to the minimum of a global fit to electroweak precision data as a 
function of the Standard Model Higgs boson mass. The yellow areas correspond to mass regions excluded 
at 95\% C.L. by direct searches at LEP and Tevatron.}
\end{figure}

\section{Experimental environment}

Higgs searches at the Tevatron depend crucially on the performance of the
accelerator and detectors. Both, CDF and D0 detectors are
currently performing close to their optimal design values, taking data
with an efficiency of about 90\%. The present Tevatron performance is
matching the design values in terms of the current weekly integrated
and peak luminosity. As of today, more than 7 fb$^{-1}$ have been
delivered to both of the experiments, while the weakly integrated luminosity 
is above 50 pb$^{-1}$ on average. If the accelerator keeps following the designed luminosity
evolution, an integrated luminosity of about 12 fb$^{-1}$ will be
achieved by the end of 2011, which increases the potential for a Higgs
discovery at the Tevatron significantly.

The two general purpose experiments at the LHC, ATLAS and CMS, have
been optimised for the discovery of a Higgs mass
up to 1 TeV. An integrated luminosity of around 10~fb$^{-1}$
will be required to cover this entire mass range and data taking is expected 
to start end of 2009.

\section{Higgs searches at the Tevatron}

\section*{Standard Model Higgs searches}
Production cross sections for the SM Higgs boson at the Tevatron are
rather small. They depend on the Higgs mass and are about 0.1 -- 1 pb
in the mass range of 100 -- 200 GeV. The largest production cross
section comes from gluon fusion, where the Higgs is produced via a
quark loop. The second largest cross section, almost an order of
magnitude smaller, is the associated production with vector bosons.
At the mass range covered by the Tevatron, below 135 GeV the highest
branching ratio is given by the decay to $b\bar b$ pairs and for
masses above 135 GeV the Higgs boson decays mainly to $WW$ pairs.

These production and decay properties lead to the following search
strategy at the Tevatron:

\begin{itemize}
\item For masses below 135 GeV the main search channels are the
associated productions with vector bosons where the Higgs decays into
$b\bar b$ pairs. In order to isolate the main background processes to
these channels, an efficient b-tagging algorithm and a good di-jet mass
resolution are essential. The same final state produced via the gluon
fusion process leads to a higher cross section but is overwhelmed by
the huge multijet QCD background at a hadron collider.
\item For masses above 135 GeV the search is mainly focused on the gluon
fusion production process where the Higgs decays into $WW$ pairs.
\end{itemize}

\begin{figure}[t] 
\label{Zmass}
  \begin{center}
    \includegraphics[height=10pc]{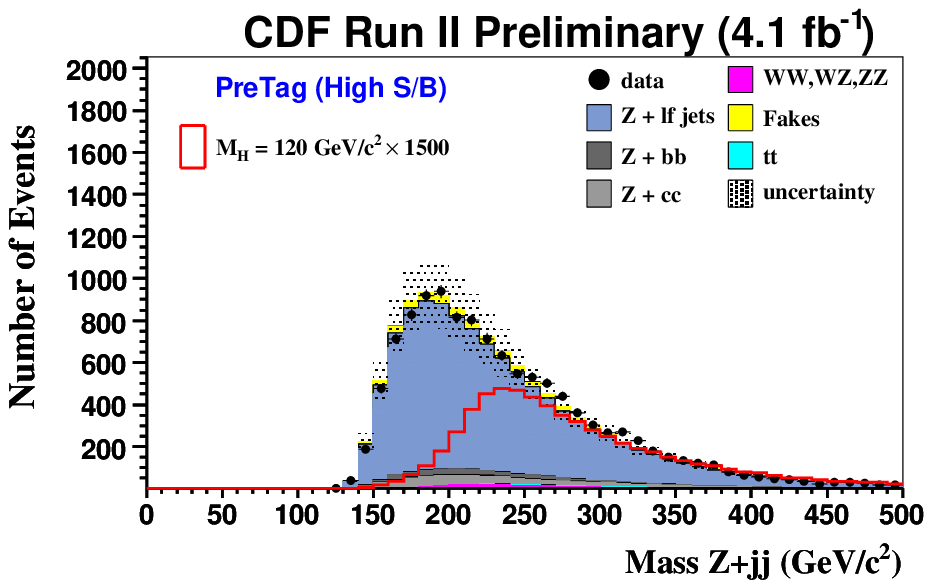}
  \caption{Mass of the $Z+$jj system in the $\ell\ell b\bar b$ final 
state at CDF prior to b tagging.}
 \vspace{1pc}
    \includegraphics[height=10pc]{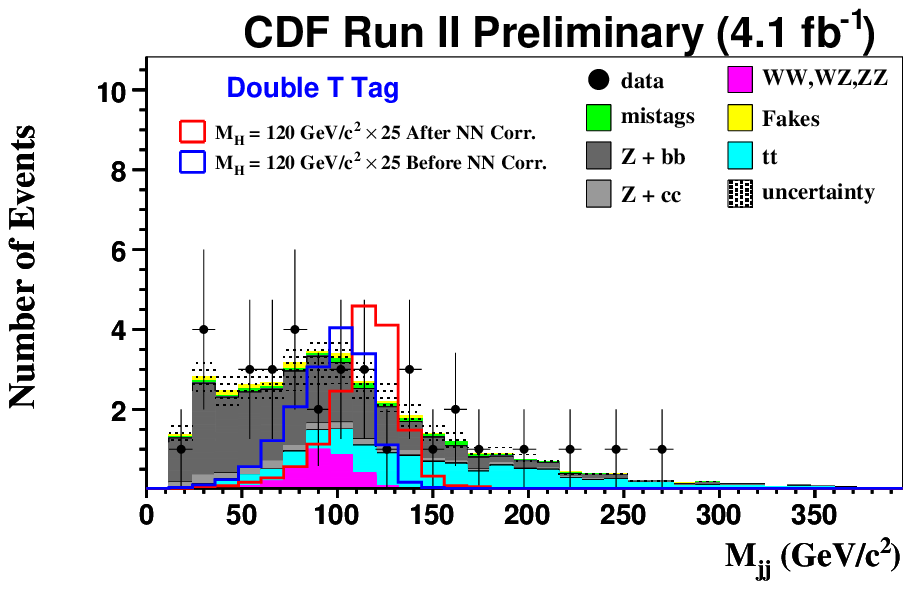}
  \caption{Di-jet mass in the $\ell\ell b\bar b$ final 
state  at CDF  after requiring two b tagged jets. The signal is plotted before (blue) and after (red) corrections 
from the 
kinematic fit.}
  \vspace{1pc}
    \includegraphics[height=10pc]{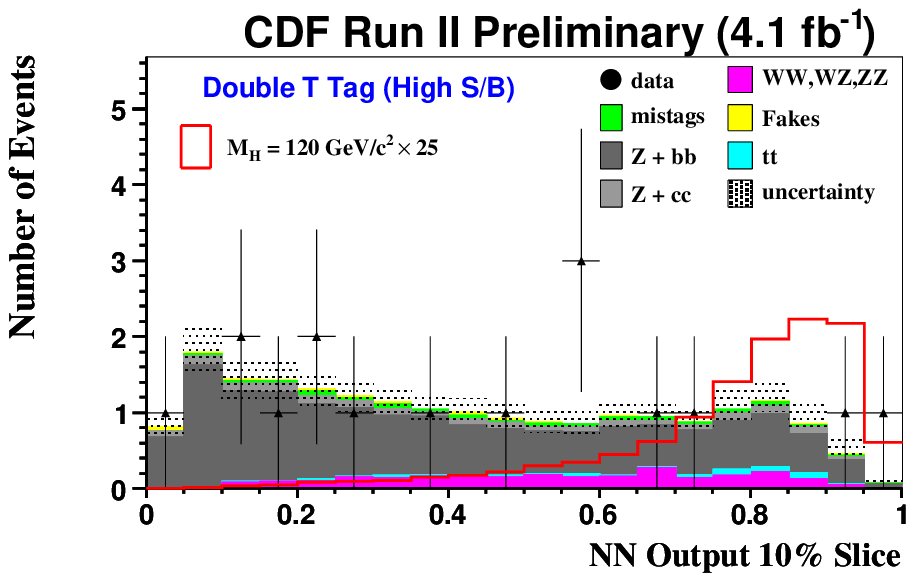}
  \caption{Projection of a two dimensional NN output, with a cut made (NNy$<0.1$) to highlight the most 
signal 
like region in the $\ell\ell b\bar b$ final state after requiring two b tagged jets.}
  \end{center}
\end{figure}

\subsection{Searches for a low mass Higgs}

Search channels of associated production with vector bosons
can be grouped into three final states. These are:
\begin{itemize}
\item no isolated lepton, missing E$_{\rm T}$ and two b jets 
\item one isolated lepton, missing E$_{\rm T}$ and two b jets
\item two isolated leptons and two b jets
\end{itemize}
Each of these final states has its advantages and disadvantages. The signal for the 
final state with one isolated lepton comes mainly from associated production of a $W$ boson 
which has the largest signal production cross section among these processes. The final state with the lowest 
background contribution is the one with two isolated leptons. It also has the advantage that it is fully 
constrained, both, the Higgs and $Z$ boson resonances can be reconstructed. The final state without any 
isolated leptons has the highest signal contribution. The $Z$ boson branching fraction into neutrinos is three 
times higher than into electrons and muons, in addition, half of the signal contribution comes from associated 
production with a $W$ boson where the lepton from the $W$ decay escapes detection. This final state has, 
however,most common large instrumental backgrounds which are difficult to handle. 

The main backgrounds to these production modes arise from two sources, physics and instrumental 
backgrounds.  Physics backgrounds are estimated from Monte Carlos (MC) and are mainly from $W$/$Z$ 
plus jets, 
diboson, top anti-top and single top production. Instrumental backgrounds are due to multijet events with 
mismeasured missing E$_{\rm T}$ or jets faking leptons. It is challenging to model these backgrounds with 
MCs and they are estimated from data sideband regions. 

Searches in all three final states can be described as a three step approach. The first step is 
to select events consistent with $W$/$Z$ and 2 jets. As an example, Fig.~2 shows the mass of the 
$Z$ plus two leptons system after the basic selection in the $\ell\ell b\bar b$ analysis \cite{llbb}.
As can be seen, the background at this stage of 
the selection is completely dominated by $Z$ plus light flavor jets events, and the signal to background ratio  
is still very small, (the signal contribution is scaled by 1500 in this plot).

The second step is to tag b jets which takes advantage of the large Higgs to $b\bar b$ branching fraction. 
b jet tagging exploits B meson lifetime, mass, fragmentation and  decay modes to separate b from light-quark 
jets. Both experiments use neural networks for optimal 
combination of tagging information. Typically the events are separated in two orthogonal categories with 
either two loosely tagged jets and 
one tightly tagged jet. Where loose tag jets are tagged with an 
efficiency of ~75\%, and tight tag jets with an efficiency of ~50\%.
The corresponding mistag rates, i.e., the probabilities to wrongly tag 
$u,d,s,g$ jets, are ~5\% and ~0.5\%, respectively. 
These values apply to jets with $p_{T}>30$~GeV and $|\eta|<0.8$ at D0, with similar tagging 
efficiencies at CDF. 

b tagging reduces the backgrounds in the basic selection by almost two orders of magnitude. 
Fig.~3 shows the 
di-jet invariant mass of the $\ell\ell b\bar b$ final state. This is the same selection as in Fig.~2
but after requiring two b jets. After b jet selection the background composition changes significantly and is 
now dominated by $W$/$Z+b\bar b$,  diboson and top anti-top production.  At this stage the most signal over 
background discriminating quantity is the di-jet invariant mass originating from the Higgs decay.

The separation power can be further optimized with the use of multivariate discrimination. Most common 
in Higgs searches at the Tevatron are techniques of Neural Network, Decision Tree and Matrix 
Element Likelihood. The basic idea for all 
of these is to exploit information from several final state variables and correlations among them. Fig.~4 
shows the output of a Neural Network for the  $\ell\ell b\bar b$ final state where the signal was trained 
against the $Z+b\bar b$ background. 

Both Tevatron experiments make large efforts to improve in all areas of these analyses, from the selection to 
the final discrimination step. An example to increase signal acceptance is the use of looser 
lepton identification criteria. Both, D0 and CDF use in the $\ell\ell b\bar b$ final 
state events with isolated tracks, electrons from less well instrumented regions of the 
calorimeter or minimal ionizing particles that escaped detection in the muon chambers.
In addition, leptonic final states now also include hadronic $\tau$ decays of 
vector bosons. 

Similar efforts are underway for a better signal to background discrimination at the final selection stage. Both 
experiments are developing further algorithms after heavy flavor  tagging to discriminate b and c quark 
jets and to discriminate a single b jet against two merged b jets. 
Since the most discriminating quantity is the di-jet 
mass, there are a lot of efforts to improve the di-jet mass resolution at both experiments. The  $\ell\ell b\bar b$ 
final state is ideal  in this respect since it is fully reconstructed and has not intrinsic missing 
E$_{\rm T}$ in the decay. 
These constraints can be effectively used to improve the di-jet mass resolution, since the detector resolution 
to the high $p_{T}$ leptons is significantly better compared to the jet resolution. Fig.~3 shows the signal mass 
peak before (blue) and after (red) this resolution improvement. This correction improved the sensitivity 
by ~10\% of 
this analysis. In addition, multivariate techniques for the final discrimination can be combined to increase 
the sensitivity further. This can be for example a matrix element likelihood as input into a final decision 
tree, or the separate training against different dominant backgrounds. The $\ell\ell b\bar b$ final state 
presented in Fig.~4 used a two dimensions neural network which is trained separately against 
the two major backgrounds, $Z+b\bar b$ and top anti-top pair production. Fig.~4  shows a slice 
from the output on the 2D network.

With an integrated luminosity of up to 5 fb$^{-1}$, 
cross section limits from individual channels are
a factor 4-8 larger than the SM prediction
at Higgs masses around 115 GeV and the combination of all contributing channels 
crucial. The systematic uncertainties for these searches are typically a total of 15\% for the signal where 
the main contributing uncertainties are theory uncertainties for the cross section calculations, 
b-tagging and object identification efficiencies. 
The total background 
uncertainty is up to twice as large (25-30\%) and is mainly due to normalisation of the $W/Z$ plus jets 
heavy flavor samples, modeling of multijet and $W/Z$ plus jets background and b-tagging.
At high discriminant values signal over background ratios are typically 1/10 - 1/20 for the 
most sensitive low mass channels. 

\subsection{Searches for a high mass Higgs}

\begin{figure}
\label{ee}
  \begin{center}
    \includegraphics[height=12pc]{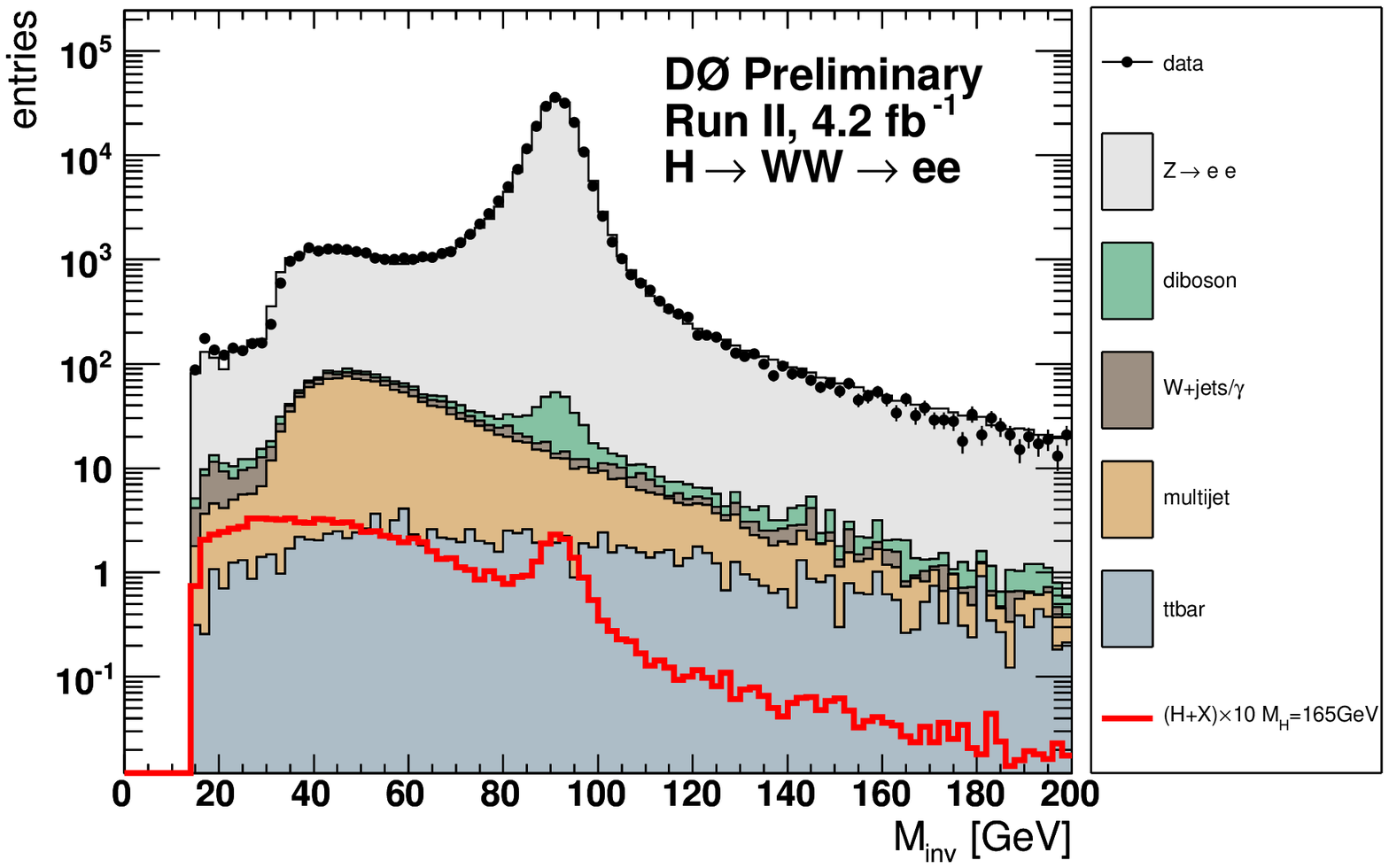}
   \vspace{1pc}
    \includegraphics[height=12pc]{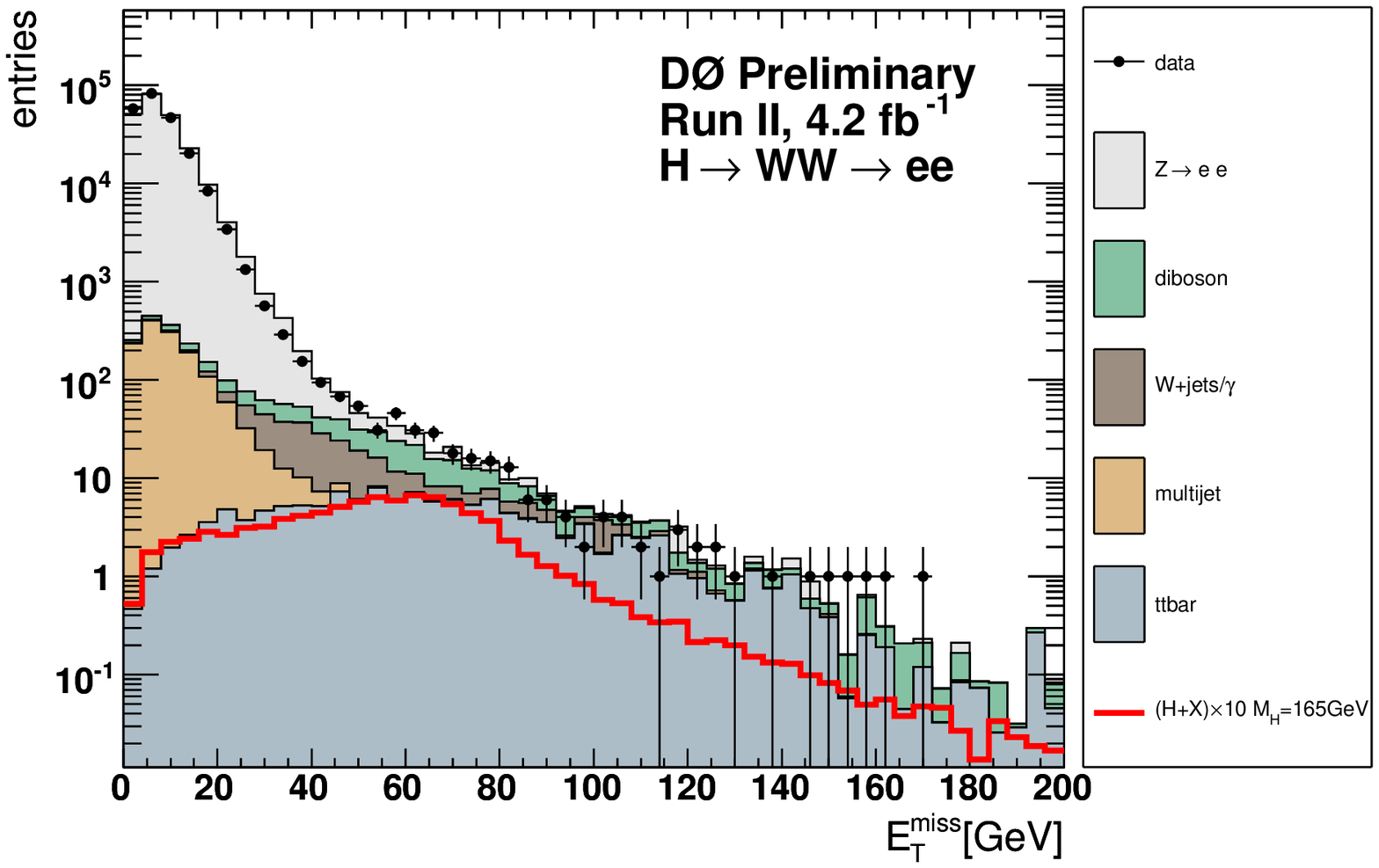}
  \caption{ Distributions of di-electron mass (top) and missing E$_{\rm T}$ (bottom) in the 
  $H\to WW^{(*)}\to e\nu e\nu$ final state at D0.} 
 \end{center}
\end{figure}

Above a Higgs mass of 135 GeV the dominant decay mode is to a pair of $W$ bosons. With the 
clean environment of the subsequent leptonic decay modes of the vector bosons one can take 
advantage 
of the dominant production mode from gluon fusion. However, there is also important signal 
contribution from 
vector boson fusion and associated production with vector bosons. For this reason the most 
sensitive search channel for a high mass Higgs boson considers all sources of opposite sign dilepton 
plus missing E$_{\rm T}$. 

The main backgrounds to this final state are Drell-Yan production, diboson, top anti-top pair 
production, $W$ plus jets and multijet production. Whereas SM $W$ pair production is an irreducible 
background. Upper Fig.~5 shows the dilepton invariant mass distribution after the basic selection \cite{hww}. 
The dominant background at this stage is Drell-Yan production, which can be reduced with cuts on 
missing E$_{\rm T}$ and missing E$_{\rm T}$ significance variables which take into account that the  
missing E$_{\rm T}$ can be caused by mismeasurement of leptons or jets. As an example lower Fig.~5 
shows the missing 
E$_{\rm T}$ distribution from D0's di-electron final state, which is plotted before the 
missing E$_{\rm T}$ cut of 20 GeV.

Spin correlation gives the main discrimination against the irreducible 
background from non-resonant SM $W$ 
pair production. In contrary to this process, in
the Higgs decay the $WW$ comes from a spin zero particle and the 
leptons prefer to point in the same direction. This feature is exploited with the use of the distribution of 
the di-lepton opening angle in the azimuthal plane. 

To increase sensitivity the D0 analysis splits the samples according to lepton flavor. A 
Neural Network with 11 kinematic and topological input variables is trained against the sum of all 
backgrounds at each hypothetical Higgs mass value in bins of 5 GeV. Such a combined evaluation of 
many kinematic variables in the Neural Network became increasingly important as the additional 
signal contributions to the gluon fusion are contributing as well. The Neural Network output of the D0
\ analysis is plotted in Fig.~6 where the signal contribution is drawn without any additional 
enhancement factor. 
 
 \begin{figure}[t]
  \begin{center}
    \includegraphics[height=11pc]{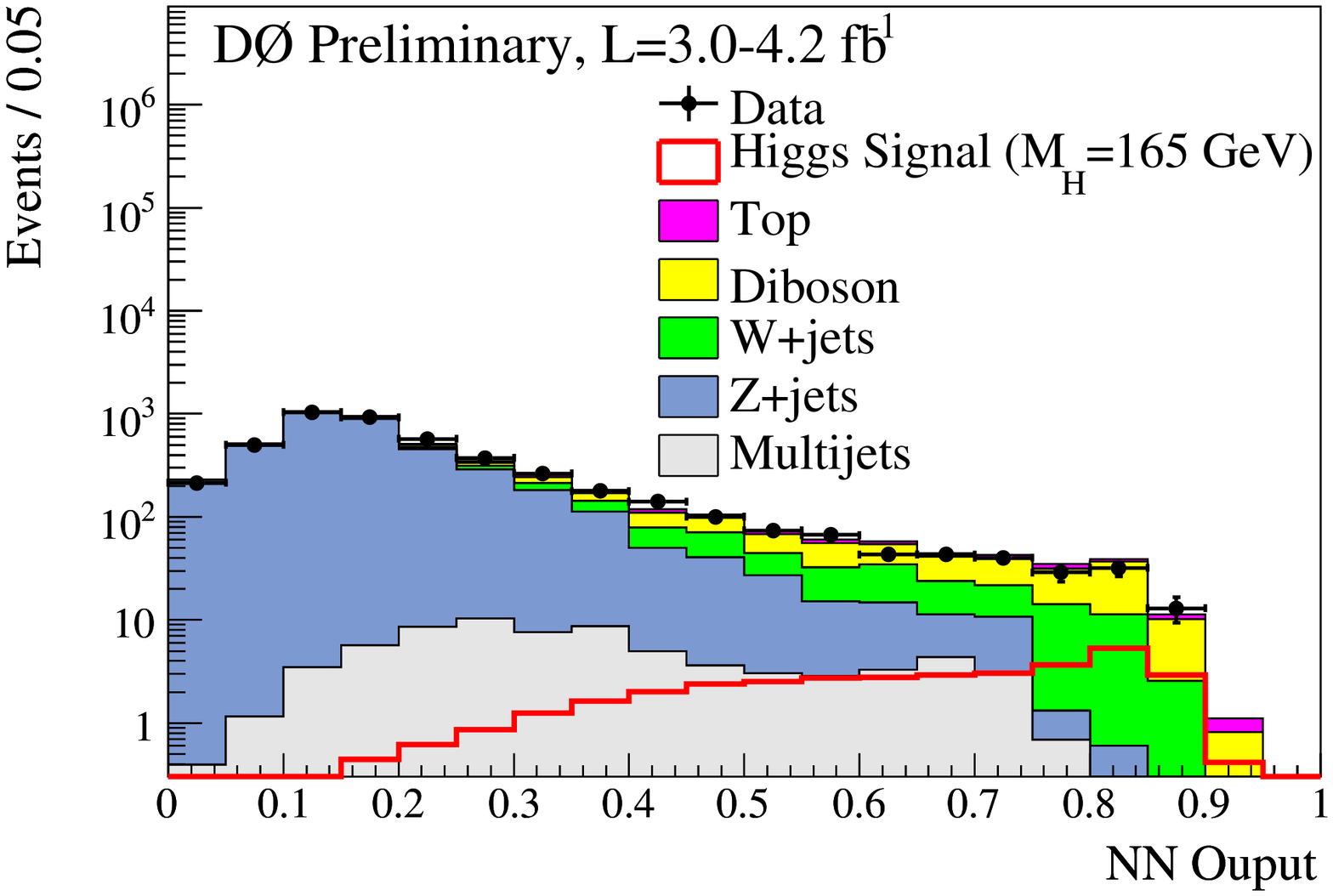}
  \caption{Neural Network output variable for the $H\to WW^{(*)}$ search at D0 for a hypothetical Higgs 
mass of 165 GeV. }
  \vspace{1pc}
    \includegraphics[height=11pc]{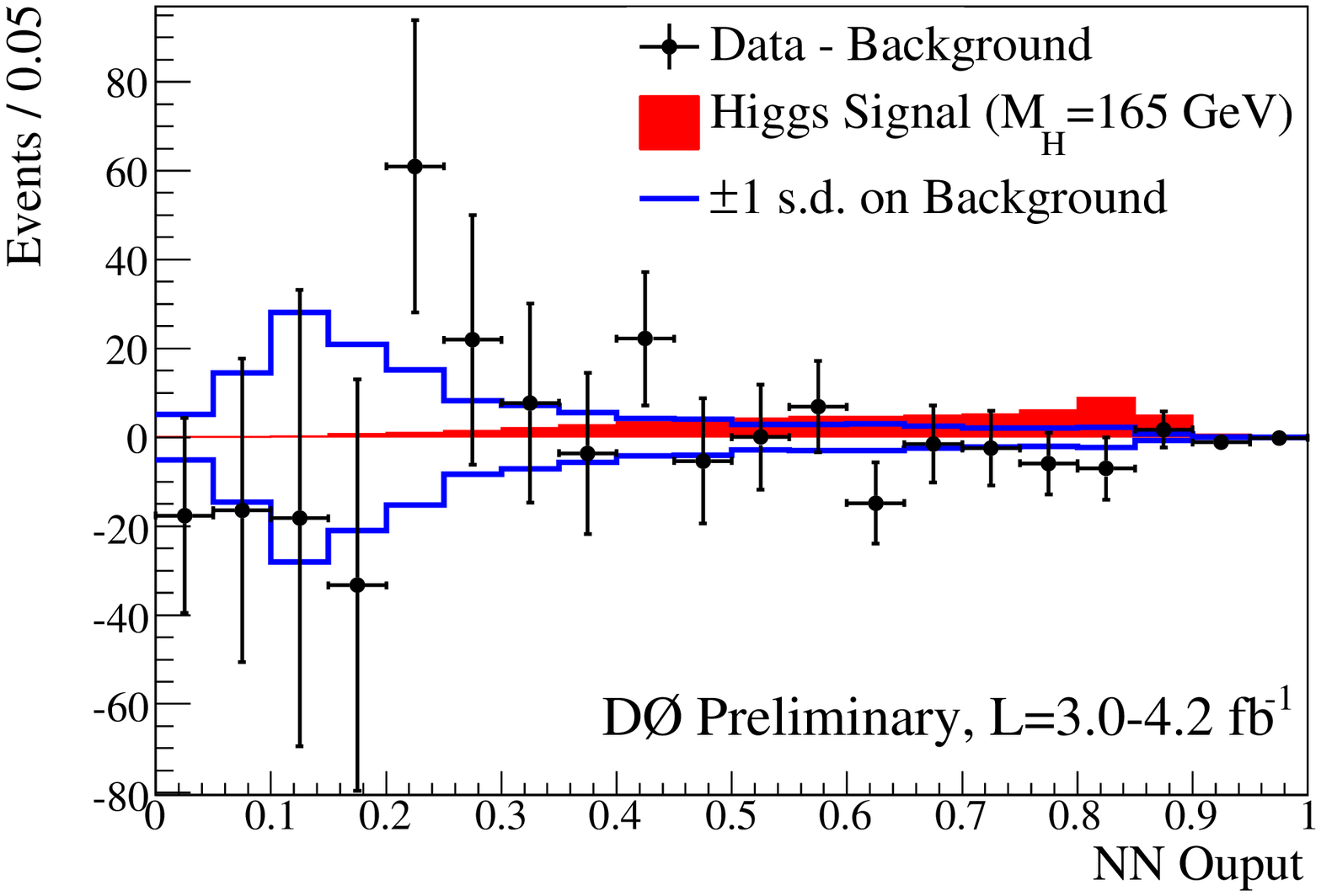}
  \end{center}
  \caption{Data and SM signal expectation after background subtraction as a function of the neural net output 
variable. The background resulting from the constrained fit using the signal plus background hypothesis is 
subtracted. The standard model signal expectation is shown by the red histogram. The constrained total 
systematic uncertainty is shown by the blue line.}
\end{figure}

The CDF analysis splits the samples into jet multiplicity and lepton identification criteria. 
In particular the separation of the sample into different  jet multiplicity bins with different 
signal and background compositions makes the final discrimination with the Neural Network effective. 
In addition, events with tight b-tagged jets are vetoed to reduce the $t\bar t $ background 
and in the Neural Network the output of a 
Matrix Element Likelihood is used in the 0-jet case.

Main systematic uncertainties for the high mass Higgs searches are from theoretical cross 
sections, lepton identification and trigger for the signal and yield 
about a 10\% error in total. For the backgrounds 
the total uncertainty is about ~15\%, coming from theory 
uncertainties for the cross section calculations, jet to lepton fake rate, 
jet identification/resolution/calibration. 
Variations due to these systematic uncertainties are propagated through 
the entire analysis such that these effect also the shape of the final discriminant. 
Fig.~7 compares the SM signal expectation (red) with 
the data after background subtraction. The constrained total systematic uncertainty is plotted in blue. 
The expected 165 GeV SM Higgs signal would be visible over the background uncertainty. The 
exclusion limits per experiment are around 1.2  -- 1.4 for the most sensitive mass region which is at 
165 GeV. At high Neural Network values signal over background ratios are close to 1.
With additional luminosity and improvements (e.g. additional channels) single experiment 
exclusion around Higgs masses of $165\pm 5$ GeV can be expected in the near future.

\begin{figure}[t]
  \begin{center}
    \includegraphics[height=16pc]{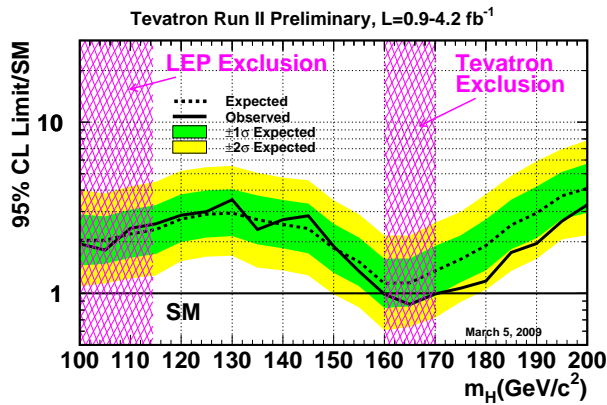}
  \end{center}
  \vspace{-1pc}
  \caption{Observed and expected 95\% C.L. upper limits on the ratios to the SM cross section, as functions of 
the Higgs boson mass for the combined CDF and D0 analyses \cite{tevcombo}.}
\end{figure}

\begin{figure}[t]
  \begin{center}
    \includegraphics[height=15pc]{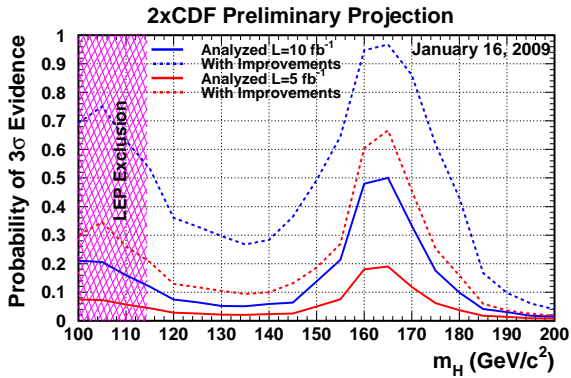}
  \end{center}
  \vspace{-1pc}
  \caption{Probability of seeing a $3\sigma$ excess at the Tevatron as a function of the Higgs mass for 
analyzed integrated luminosities of 5 fb$^{-1}$ and 10 fb$^{-1}$ per experiment, with the current 
performance and with expected improvements by the end of the Tevatron run.}
\end{figure}

\subsection{Combined Standard Model Higgs limits}

In order to reach the highest sensitivity a full combination of all channels from CDF and D0 is 
performed \cite{tevcombo}. In such a Tevatron combination 75 different orthogonal 
channels are considered with more 
than 50 different systematic uncertainties taking into account correlations between channels and 
across experiments. Both a bayesian and a modified frequentist approach have been used for the
combination, and results have been shown to agree within 10\%.

Fig.~8 shows the combined cross section limit relative to the SM expectation. The limits are 
expressed as a multiple of the SM prediction for test masses (every 5 GeV) for which both experiments 
have performed dedicated searches in different channels. The points are joined by straight lines for better 
readability. The bands indicate the 68\% and 95\% probability regions where the limits can fluctuate, in the 
absence of signal. The limits displayed in this figure are obtained with the Bayesian calculation.
This result is based on an effective luminosity of 2.6 fb$^{-1}$ around masses of 115 
GeV and 3.8 fb$^{-1}$ at masses around 160 GeV. 
Observed and expected limits agree within one standard deviation 
and no indication of a Higgs boson signal has been observed. A mass range of 160 to 170 GeV has 
been excluded at the 95\% C.L. This is the first direct exclusion of a SM Higgs in a mass range 
above the LEP limits. 

The Tevatron collider is expected to deliver an integrated luminosity of 12 fb$^{-1}$ per experiment by the 
end of 2011. This corresponds to a final dataset of 10 fb$^{-1}$ per experiment after accounting
for data taking efficiencies. 
Fig.~9 shows the probability of seeing a $3\sigma$ excess as a function of the Higgs mass for 
analyzed integrated luminosities of 5 fb$^{-1}$ and 10 fb$^{-1}$ per experiment, assuming CDF and D0 
perform the same \cite{future}. 
Two scenarios are shown, in which channels have the same performance as for the 
Winter 2009 combination (solid lines), and for the case with another factor of 1.5 increase in sensitivity 
(dashed lines). With expected future improvements in the analyses there can be a 50\% chance to find 
evidence for a 115 GeV Higgs boson should it exist.

\section*{MSSM Higgs searches}

\begin{figure}
  \begin{center}
    \includegraphics[height=14pc]{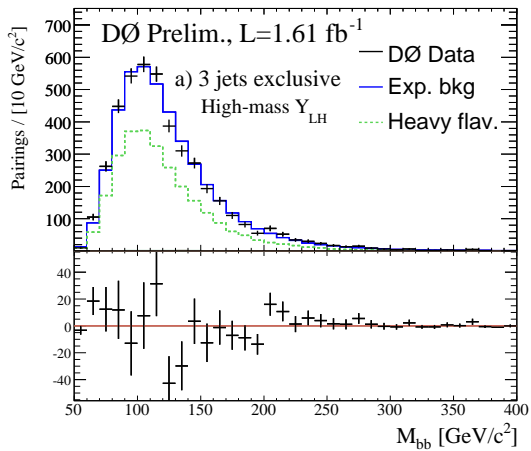}
  \end{center}
  \vspace{-1pc}
  \caption{Invariant mass for the exclusive three-jet channel of the $b \phi\to bbb$ search at D0. The solid 
blue line shows the total background estimate, and the dotted green line represents the heavy flavor 
component. The lower panels show the difference between the data and the background expectation.}
\end{figure}

The Minimal Supersymmetric Standard Model (MSSM) predicts two Higgs
doublets leading to five Higgs bosons: a pair of charged Higgs boson
($H^\pm$); two neutral CP-even Higgs bosons ($h$,$H$) and a CP-odd Higgs
boson ($A$). At tree level, the Higgs sector of the MSSM is fully
described by two parameters, which are chosen to be the mass of the
CP-odd Higgs, $m_A$, and $\tan\beta$ , the ratio of the vacuum
expectation values of the two Higgs doublets. The Higgs production
cross-section is enhanced in the region of low $m_A$ and high
$\tan\beta$ due to the enhanced Higgs coupling to down-type fermions.
This leads to a significant increase in production cross section of neutral 
Higgs bosons $\phi$ in gluon-fusion (via a b quark loop) as well as in associated production with b quarks. 
This makes it possible to search in the MSSM for the inclusive productions:
$\phi\to\tau\tau$, $b \phi\to b\tau\tau$, $b \phi\to bbb$, 
which would be very challenging in the SM due to the smallness 
of the cross section and the large
irreducible backgrounds of $Z\to\tau\tau$ and multijet production. In the low $m_A$, high
$\tan\beta$ region of the parameter space, Tevatron searches can
therefore probe several MSSM benchmark scenarios extending the search
regions covered by LEP \cite{lepmssm}.

Both collaborations analyzed these channels, searching for a resonance n the di-tau or 
di-b-jet mass spectrum. 
Fig.~10 shows the invariant mass distribution for the 3 jets channel in the $b \phi\to bbb$ search. 
No significant excess has been observed in any of the channels, allowing to set limits on production of 
neutral Higgs bosons. 

The results on these three search channels at D0 were combined to obtain upper limits on $\tan\beta$ as a 
function of $m_{A}$ \cite{d0mssmcombo}.
Radiative corrections to the Higgs couplings introduce a dependence on other model 
parameters of the MSSM so the combined result is provided within various benchmark models using the 
same combination 
technique as has been employed for the combination of SM Higgs boson searches. Fig.~ 11 shows 
the region excluded in the ($m_{A}$, $\tan\beta$)-plane at 95\% C.L. within the Ò$m_{h}$-maxÓ scenario, 
based on an integrated luminosity of 1.0 - 2.6 fb$^{-1}$. A similar combination across the two 
experiments was done for the $\phi\to\tau\tau$ results from CDF and D0, which has a similar exclusion 
region in the parameter space \cite{tevmssmcombo}. It is expected that with the full Run II dataset and 
the combination of all MSSM channels across the two experiments sensitivity down to $\tan \beta$ values of 
20 can be reached in the different benchmark scenarios.

\begin{figure}
  \begin{center}
    \includegraphics[height=15pc]{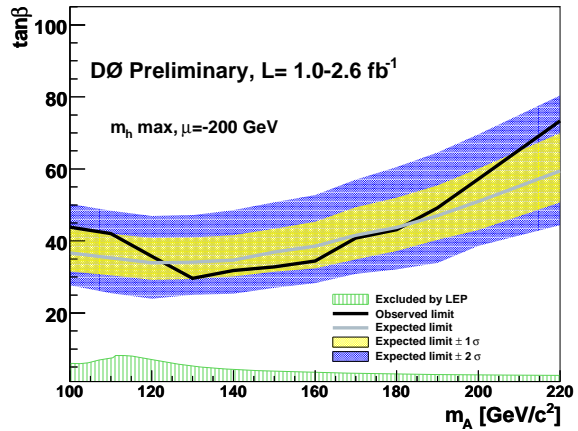}
  \end{center}
  \vspace{-1pc}
  \caption{ 95\% confidence limits in the $(m_{A},\tan\beta)$ plane for the maximal mixing with $\mu < 0$ 
benchmark scenario. The dark line shows the observed limit, the gray line the expected limit and 
the shaded light-green area shows the limits from LEP.}
\end{figure}

\section{Higgs searches at the LHC}

With a centre of mass energy of 14 TeV
and a high design luminosity of $10^{34}$~cm$^{-2}$~s$^{-1}$ the LHC offers the best opportunity to make 
a direct observation of the Higgs boson, and both general purpose experiments, ATLAS and CMS will search 
for its existence. 

Higgs production cross sections at the LHC are roughly two orders of magnitude larger than at the 
Tevatron and the dominant production channel is gluon fusion. This is followed by
vector boson fusion production which plays at the LHC an important contribution. From Tevatron to LHC 
energies the relevant backgrounds, like top anti-top pair production, rise often faster then the signal. 
The main handle against the huge multijet backgrounds at the LHC will be mainly to search in the leptonic 
decay modes.

The main discovery channels at the LHC are the inclusive searches $H\to\gamma\gamma$ and $H\to WW$ 
or $H\to ZZ$. For a low mass Higgs discovery, $H\to\gamma\gamma$ is one of the most promising channels. It has a 
small branching ratio, thus it needs good calorimeter resolution
to observe a narrow mass peak in the 
prompt $\gamma\gamma$ continuum. The inclusive search for $H\to WW$ is similar to the same search at 
the Tevatron and has it highest sensitivity in the same mass region around 160 GeV.  
$H\to ZZ$ with the decay 
into four leptons is called the ``gold plated decay channel'' due to the very little background contribution 
especially for Higgs masses where one of the $Z$ bosons has to be off-shell. 

These channels will be complemented with more exclusive searches. In particular,  the vector boson fusion 
(VBF) production will provide additional sensitivity. In the VBF process the two incoming quarks of the 
colliding protons radiate two vector bosons and the Higgs is produced via the fusion of these 
two electroweak bosons. 
These
quarks emerge as low angle `tag' jets which can be observed. The lack
of color exchange between the tag jets leads to a suppression of
radiation between them and thus to a lack of hadronic activity in the
central region of the detector. These two features are in contrast
with most of the background processes and allow the signal to be 
enhanced significantly. 

Fig.~12 shows as an example the ATLAS discovery potential \cite{atlas}, a
similar sensitivity is expected from CMS. This 
plot shows the luminosity required for a $5\sigma$ discovery with the combination 
of the main channels with 14 
TeV of collider energy. There are basically two main scenarios within the mass range preferred by the 
electroweak fits. If the Higgs boson is heavier than 130 GeV discovery with a few fb$^{-1}$ will be already 
possible.  If the Higgs is lighter than 130 GeV, the luminosity required for its discovery increases rapidly. A 
discovery of a light Higgs would be more challenging and a few years of running may be needed. 
The LHC experiments will also cover a broad program of searches for Higgs bosons beyond the SM, 
where the Tevatron searches will be extended to significantly higher Higgs boson masses.

\begin{figure}[t]
  \begin{center}
    \includegraphics[height=16pc]{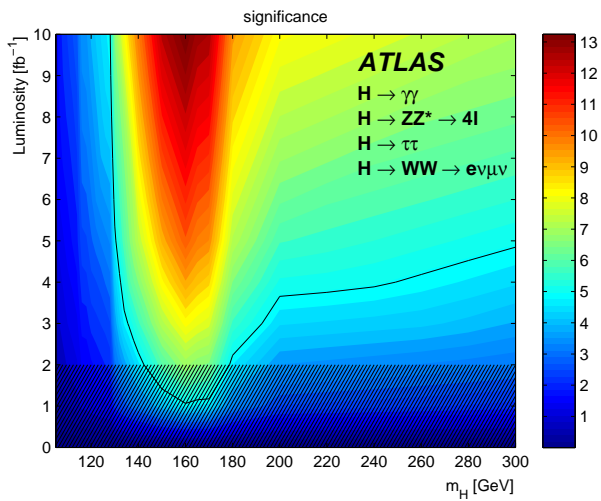}
  \end{center}
  \vspace{-1pc}
  \caption{Significance contours for different Standard Model Higgs masses and integrated luminosities as 
expected at the ATLAS experiment. The 
thick curve represents the $5\sigma$ discovery contour. The hatched area below 2 fb$^{-1}$ indicates the 
region where the approximations used in the combination are not accurate, although they are expected to be 
conservative.}
\end{figure}

\section{Outlook}

With Tevatron's
excellent performance matching the designed delivered weekly
luminosities, a significant amount of sensitivity will be gained with an
increase of the luminosity by the end of 2011. This will make it possible to use 
four times more data than was used in the presented results, to search for the Higgs at 
lower masses, with the possibility of a $3\sigma$ discovery 
of a light SM Higgs boson. The reach of sensitivity around 165 GeV will be significantly 
expanded as well with the 
final dataset. To improve the di-jet mass resolution, b-tagging, multivariate techniques and to increase the
signal acceptance are the most important challenges for future Higgs searches at the Tevatron.

The LHC will offer the best possibility to search for a SM or beyond-SM Higgs bosons and with a data 
collected after a few years of running discovery is guaranteed if the Higgs boson exists and has a mass
below 1 TeV. 

\section*{Acknowledgments}
I would like to thank my colleagues from the CDF,  D0, ATLAS and CMS Collaborations
working on this exiting topic and for providing material for this
talk. I also like to thank the organizers of Physics in Collision for
a stimulating conference.



\end{document}